# Anomalous valley polarization in monolayer MoSe$_2$

Anmin Zhang,[1] Jiahe Fan,[1] Yusheng Li,[1] Jianting Ji,[1] Guihua Zhao,[1] Tianlong Xia,[1] Tengfei Yan,[2] Xinhui Zhang,[2] Wei Zhang,[1] Xiaoqun Wang[1] and Qingming Zhang[1*]

[1]Department of Physics, Beijing Key Laboratory of Opto-electronic Functional Materials & Micro-nano Devices, Renmin University of China, Beijing 100872, P. R. China

[2]State Key Laboratory of Superlattices and Microstructures, Institute of Semiconductors Chinese Academy of Sciences, P.O. Box 912, Beijing 100083, P. R. China

**Modern electronic devices heavily rely on the accurate control of charge and spin of electrons. The emergence of controllable valley degree of freedom brings new possibilities and presents a promising prospect towards valleytronics. Recently, valley excitation selected by chiral optical pumping has been observed in monolayer MoS$_2$[1-3]. In this work, we report polarized photoluminescence (PL) measurements for monolayer MoSe$_2$, another member of the family of transition-metal-dichalcogenides (MX$_2$), and observe drastic difference from the outcomes of MoS$_2$. In particular, we identify a valley polarization (VP) up to 70% for B exciton, while that for A exciton is less than 3%. Besides, we also find a small but finite negative VP for A$^-$ trion. These results reveal several new intra- and inter-valley scattering processes which significantly affect valley polarization, hence provide new insights into exciton physics in monolayer MX$_2$ and possible valleytronic applications.**

[*] qmzhang@ruc.edu.cn

The group VI transition-metal dichalcogenides $MX_2$ (M=Mo, W, X=S, Se, Te) have caught intensive attention in the past few years, mainly due to the existence of valley degrees of freedom, and most importantly, the possibility to populate electrons selectively in different valleys using optical methods in monolayer samples[4]. In comparison to bulk materials with an indirect gap, monolayer $MX_2$ features a direct band gap with two degenerate minima (valleys) located at the corners of the hexagonal Brillouin zone K and K'[4-6]. Besides, since the inversion symmetry is explicitly broken, the two valleys can be distinguished and labeled by opposite helicity. Specifically, electron hopping at K' point is allowed with right-circular optical pumping but prohibited with left-circular pumping, while the situation is reversed at K point[7, 8]. This intriguing feature thus provides us the ability to tune valley polarization (VP) using circularly polarized light[1-3].

Recently, VP has been reported in monolayer $MoS_2$[1-3] and $WSe_2$[9]. In particular, a nearly fully polarized photoluminescence (PL) spectrum has been observed for the A exciton[2], demonstrating a successful manipulation of valley degree of freedom in such materials. In addition to the well-studied A exciton, there exists another B exciton which is shifted from the A exciton by strong spin-orbit coupling[5,6]. The splitting energies are about 160 meV for $MoS_2$, 180 meV for $MoSe_2$, 430 meV for $WS_2$ and 460 meV for $WSe_2$[4,10,11]. Although a tiny PL intensity difference between left- and right-excitations at the high-energy wings of A-exciton in mononlayer $MoS_2$ was considered as a hint to the B-exciton VP [2], an unambiguous demonstration has not been witnessed yet.

Here we reported a study of polarized PL spectra for different excitons in monolayer $MoSe_2$ at various temperatures. Despite the structural similarity of group VI transition-metal dichalcogenides, the A-exciton VP of $MoSe_2$ is almost zero under both off- and near-resonance conditions. For instead, a VP as high as ~70% can be identified for B exciton under circularly polarized optical pumping. This anomalous PL polarization can be understood quantitatively by taking the inter-valley scattering and the lifetime of various excitons into account. Besides, we also observe a small but distinguishable negative polarization effect (~-4%) associated with the spectrum of A⁻ trion, which is formed by an A exciton binding with another hole or electron (A⁻, Fig. 1)[12-14], at a temperature below 50 K and under off-resonance conditions. This observation indicates a negative VP of A⁻ trions and can be attributed to inter-valley scattering of B excitons. Our findings suggest that several particular intra- and inter-valley scattering processes can have significant impact on the control of value degrees of freedom in $MX_2$ materials, hence have to be included into consideration in possible valley-based electronic and optoelectronic applications.

The PL spectra of A/B excitons and A⁻ trion measured at various temperatures are depicted in Fig. 1, all showing clear peaks at temperature below 200 K. The temperature dependence of excitonic energy follows the Varshini relationship[15] (insets of Fig. 1), which leads to zero-temperature extrapolated values of 1.640 eV for A exciton, 1.612 eV for A⁻ trion, and 1.846 eV for B exciton. Based on this measurement, the energy difference between A and B excitons is determined to be 206 meV, very close to the value of 180 meV obtained by angle-resolved photoemission and absorption spectra[16].

We now turn to the investigation of valley exciton polarization. Figure 2 shows the polarization-resolved PL spectra for A and B excitons collected under both the near- and off-resonance conditions at a temperature of 70 K. The most striking feature of these results is the significant polarization effect associated with B exciton when subjected to near-resonance optical pumping (1.96 eV). The intensity of observed PL polarization, defined as $P = (I_{\sigma+} - I_{\sigma-})/(I_{\sigma+} + I_{\sigma-})$ with $I_\sigma$ integrated intensity of PL peak, is about 70% as depicted in Fig. 2(c). This observation indicates that the B excitons of monolayer $MoSe_2$ in a given valley can be selectively pumped via optical method, leading to large valley polarization for B excitons. This polarized effect disappears under off-resonance conditions (2.33 eV), as one would naturally expect (See Fig. 2(d)).

Another interesting finding of Fig. 2 is a very tiny VP for A excitons, in drastic contrast to the case of monolayer $MoS_2$ where the polarization of PL spectrum can be nearly 100%[2]. Indeed, as shown in Figs. 2(a) and 2(b), the intensity of polarization for A exciton within monolayer MoSe$_2$ is less than 3% under both near- and off-resonance conditions. This observation has also been confirmed when a left-circularly polarized pumping laser is used (see Fig. S2 in Supplementary information). The electron/hole hopping process is illustrated in the inset of Fig. 2(a). An A-exciton at K' point excited by right-circular light can be scattered to K point via inter-valley scattering process, and emits a left-circularly polarized photon through excitonic recombination.

The depolarization of A exciton PL spectrum in monolayer MoSe$_2$ can be explained by the long life time of A excitons. Theoretically, VP can be expressed as[2]

$$P = \frac{1}{1 + 2 \times \frac{\tau_E}{\tau_V}} \qquad (1)$$

where $\tau_E$ and $\tau_V$ are exciton and valley lifetimes, respectively. It is then naturally expected that a longer exciton lifetime $\tau_E$ is detrimental to the polarization effect. For monolayer MoSe$_2$, both PL and transient reflection measurements suggest a lifetime of >125ps, which is more than an order of magnitude longer than that of monolayer MoS$_2$ (several ps)[20]. The microscopic origin of the long lifetime of A excitons may be attributed to the following factors. First, the defect concentration in monolayer MoSe$_2$ samples is usually lower than that in MoS$_2$, since creating selenium vacancies is relatively difficult due to the its large radius. The lack of defects thus leads to a longer exciton lifetime. Second, it is suggested by experiments[20] that the exciton decay process in MoSe$_2$ is dominated by the nonlinear exciton-exciton annihilation, while a linear exciton decay channel is expected for MoS$_2$. Third, the phonon spectra obtained by both experimental measurements and numerical simulation are quite different for MoSe$_2$ and MoS$_2$. That may cause different electron-phonon interactions and hence affect the exciton lifetime. Besides, it is also possible that the coupling between monolayer sample and substrate is different for each member of the MX$_2$ family.

The prominent distinction of VP for A and B excitons can also be explained via the same argument of excitonic lifetimes. Notice that the valley depolarization is dominated by electron-hole exchange interaction, which originates from Coulomb interaction between electrons/holes at different valley points and hence is determined by the orbital features of the involved energy bands. Considering the fact that A/B excitons come from the spin-split upper/lower bands having the same orbital components, and the splitting energy is small (0.4 eV) compared to the band gap (~1.9 eV), a similar electron-hole exchange interaction can be expected for both types of excitons, which in-turn leads to a comparable valley lifetime for A and B excitons.[17] Thus, the ratio between A and B exciton lifetimes can then be extracted as $\tau_A/\tau_B \sim 75$ from our measurement of VP.

An alternative route to obtain the lifetime ratio $\tau_A/\tau_B$ is by investigating the absolute PL intensity. Since the only decay channel of A exciton is via excitonic recombination, the PL intensity of A exciton is proportional to the decay rate $n_A/\tau_A$, where $n_A$ is the density of A excitons and $\tau_A$ is the corresponding lifetime. Under the steady-state condition, the decay rate is balanced by the generation rate $g_A = n_A/\tau_A$. For B exciton, however, an additional annihilation process is present through which the holes generated by photoexcitation in lower valence band move to the upper valence band[18], and cause the formation of so-called dark excitons. The lifetime of B excitons hence is determined cooperatively by the recombination rate ($1/\tau_B^*$) and the hole hopping rate ($1/\tau_D$), leading to $1/\tau_B = 1/\tau_B^* + 1/\tau_D$. In a steady state, the generation rate and decay rate are both given by $g_B = n_B/\tau_B = n_B/\tau_B^* + n_B/\tau_D$, where $n_B$ is the density of B excitons. The PL intensity, on the other hand, is only associated with the recombination channel, and is proportional to $\tau_A/\tau_B^*$. Considering the fact that both measurements[14] and calculations[19] suggest an equal generation rate for both A and B excitons, the PL intensity ratio of ~98 obtained from our experiment suggests an excitonic lifetime ratio $\tau_A/\tau_B \sim 98$, in agreement with the value of ~75 obtained from polarization measurements. The consistency indicates that the difference in excitonic lifetime is the primary factor behind the anomalous VP in monolayer MoSe$_2$. In fact, a much longer lifetime of A excitons in MoSe$_2$ than in MoS$_2$ has been recently reported[20].

Figure 3 shows the temperature dependence of polarized PL spectra of A/B excitons and A$^-$ trion. We notice that with decreasing temperature, the VP for B exciton is significantly enhanced, while that for A exciton remains imperceptible down to the lowest temperature of 10 K. Meanwhile, we identify a slight but noticeable difference between right and left-circular PL spectra of A$^-$ trion for T < 50 K, which reaches -4% at our lowest attainable temperature T = 10K. This temperature dependence indicates that the observation reported here reveals intrinsic properties of the samples, as the optical system is always kept at room temperature.

The observation of a negative VP for A$^-$ trions is quite surprising because theoretically no chiral optical valley selectivity is expected under non-resonance conditions. It must originate from exciton scattering. While the inter-valley scattering of A excitons between K and K' points is balanced as indicated by the results of zero VP, the scattering processes of B excitons can be directionally discriminated between K and K' points. If a B exciton is created at K' point by right-circular polarized pumping light, it has three scattering channels. The first one is the inter-valley scattering from K' to K points. This process would decrease the intensity of VP for B excitons, but have no effect on the PL of A$^-$ trion. The second one is the formation of dark excitons at K' points as mentioned above, which is invisible and hence does not affect the trion polarization. The third process is the hole hopping from the lower valence band at K' point to the upper valence band at K point (inset of Fig. 3(a))[18], which is highly possible because spin momentum is conserved and the hopping is energetically favorable. The net result of this process is an increase the densities of A exciton and A$^-$ trion at K point, which consequently leads to a negative VP to both types of excitations.

The observation of polarized trions rather than excitons can also be understood by examining the ratio of trion valley lifetime to its lifetime. In comparison to excitons, both theories and experiments suggest that a trion has a much longer valley lifetime. In fact, trions suffer much weaker intervalley scattering because spin flip, required by intervalley scattering process, is energetically and spin forbidden [17,28,29]. For example, the trion valley lifetime is ~ns in WSe$_2$, while the exciton valley lifetime is only several ps [30]. This extremely long valley lifetime of trions thus favors the observation of the associated polarization effect in WSe$_2$, provided that the appropriate light sources which can excite the B excitons completely become available, and high signal-to-noise ratio is attainable at low enough temperatures.

The temperature-dependent intensities of VP for A/B excitons and A$^-$ trion are summarized in Fig. 4. The VP of B exciton approaches a saturation value of 70% below 80 K, with a lost of 30% possibly induced by defects or substrates[21-24]. The temperature evolution can be well described by a phenomenological model of phonon-dominated inter-valley scattering, where the intensity of VP is expressed as

$$P = \frac{N_0 + a \times (N_1 + N_2) - (1-a) \times (N_1 + N_2)}{N_0 + a \times (N_1 + N_2) + (1-a) \times (N_1 + N_2)} = 1 + \frac{(2a-2) \times (N_1 + N_2)}{N} \quad (2)$$

Here, we define $N$, $N_0$, $N_1$ and $N_2$ as the total number of B excitons, the number of the excitons staying at K' points, the number of the excitons scattered to K points by temperature-independent factors like defects, and the number of the excitons scattered to K points by phonons, respectively, and $a$ is the scattering probability between K and K' points. We further assume that both $a$ and $N_2$ are proportional to phonon number $n = 1/(\exp(\hbar\omega/k_B T) - 1)$ with $k_B$ Boltzmann constant, and $\omega$ the frequency of phonon branch at K point which dominates electron-phonon coupling. The fitting using equation (2) with the number constraint $N_0 = N - N_1 - N_2$ quantitatively re-produces the experimental results (green curve in Fig. 4), and gives a phonon frequency of 186 cm$^{-1}$, which is in perfect agreement with the calculations[25].

In summary, we have systematically studied circularly-polarized PL spectra in monolayer MoSe$_2$. In contrast to the case of MoS$_2$ and WSe$_2$, the VP of A exciton is very small (<3%) under both near- and off-resonance conditions, while a significant polarization as high as 70% is observed for B exciton at low temperatures. The anomalous behavior of polarized PL spectra is quantitatively understood within a phenomenological model incorporating different lifetimes of A and B excitons. In addition, we also observe a small but identifiable negative VP for A$^-$ trion at temperatures below 50 K, which is attributed to inter-valley scattering of B excitons. Our results suggest that the control of valley degrees of freedom in MX$_2$ systems can be drastically affected by some intra- and inter-valley scattering processes. The study also demonstrates that B exciton is very insensitive to defect scattering in comparison to A exciton. This offers an alternative candidate for valleytronics applications and allows a more flexible valleytronics device design.

## Methods

**Sample preparation.** MoSe$_2$ crystals were synthesized by chemical vapor transport method as described in Ref. 26. Monolayer samples were obtained from bulk crystals via mechanical exfoliation, and then attached on silicon substrate with a 300 nm SiO$_2$ buffer layer. Atomically thin MoSe$_2$ was first visually identified with interference color through an optical microscope, as depicted in the inset (a) of Fig. S1. The nearly trapezoidal monolayer MoSe$_2$ flake covers an area of 10 μm x 5 μm and is well isolated from other pieces, hence preventing possible complication of multilayer signals. The thickness was further confirmed by frequency measurement of A$_1$' Raman mode at 241.1 cm$^{-1}$ (inset (d) of Fig. S1)[27]. The strong PL peak associated with A excitons provides another evidence for the thickness of our samples, as the PL spectra show a dramatic drop of intensity in bilayer and multilayer cases due to the transition from a direct to an indirect gap. An AFM image is provided in supplementary information (Fig, S1) to clearly demonstrate the thickness of MoSe$_2$.

**PL measurements.** The PL measurements with laser sources of 633 nm (Melles Griot, HeNe Laser) and 532 nm (Torus 532, Laser Quantum) were carried out using Jobin Yvon T64000 spectrometer. The measurements with 725 nm (fianium sc450) were performed using Jobin Yvon iHR550 monochromator, with a band-pass filter (Andover TBP01-704, centered at 747 nm and bandwidth 33 nm). The laser spot size was typically ~5μm. The laser power was stabilized at a level of 100 μW and carefully monitored with a power meter (Coherent Inc.), in order to avoid overheating and ensure accurate PL measurements. In the context, the pumping light is right-circularly polarized ($\sigma^+$), unless specified explicitly.

## Ackonwledgements

This work was supported by the Ministry of Science and Technology of China (973 projects:2012CB921701 and 2011CBA00112 ) and the NSF of China (Grant No.: 11034012 & 11174367). Q.M.Z. was supported by the Fundamental Research Funds for the Central Universities, and the Research Funds of Renmin University of China. T.-L. X. was supported by NSFC (Grant No. 11004245) and the Fundamental Research Funds for the Central Universities, and the Research Funds of Renmin University of China (Grant No. 14XNLQ07).

Figure captions:

**Fig. 1** PL spectra of A excitons and A$^-$ trions (a) and B excitons (b) in monolayer MoSe$_2$ with 633 nm excitation at selected temperatures. Temperature dependence of PL peak positions is shown in the insets, in which the red fitting curves follow Varshni empirical relationship. The sharp peaks and the features superimposed on PL peak in (b) are first- and higher-order phonon modes enhanced by resonance excitation.

**Fig. 2** Polarized PL spectra with different excitation energies at 70 K and the polarization of excitation light is $\sigma^+$. $\sigma^+/\sigma^-$ denotes right/left-circular polarization. Excitation energies: (a) 1.71 eV, near-resonance for A excitons. A$^-$ peak is cut off by band-pass filter; (b) 1.96 eV, off-resonance for A excitons and A$^-$ trions; (c) 1.96 eV, near-resonance for B excitons; (d) 2.33 eV, off-resonance for B excitons. The insets in (a)&(b) and (c)&(d), are hopping processes under the corresponding excitation energies for A and B excitons respectively.

**Fig. 3** Polarized PL spectra with 633 nm (1.96 eV) excitation at selected temperatures. (a) Temperature evolution for A$^-$ trions and A excitons. The inset illustrates a hole scattering from lower valence band at K' point to upper valence band at K point; (b) Temperature evolution for B excitons.

**Fig. 4** Temperature dependence of VP for A/B excitons under resonant conditions and A$^-$ trions under non-resonant conditions. The green curve is the fitting result using equation (2).

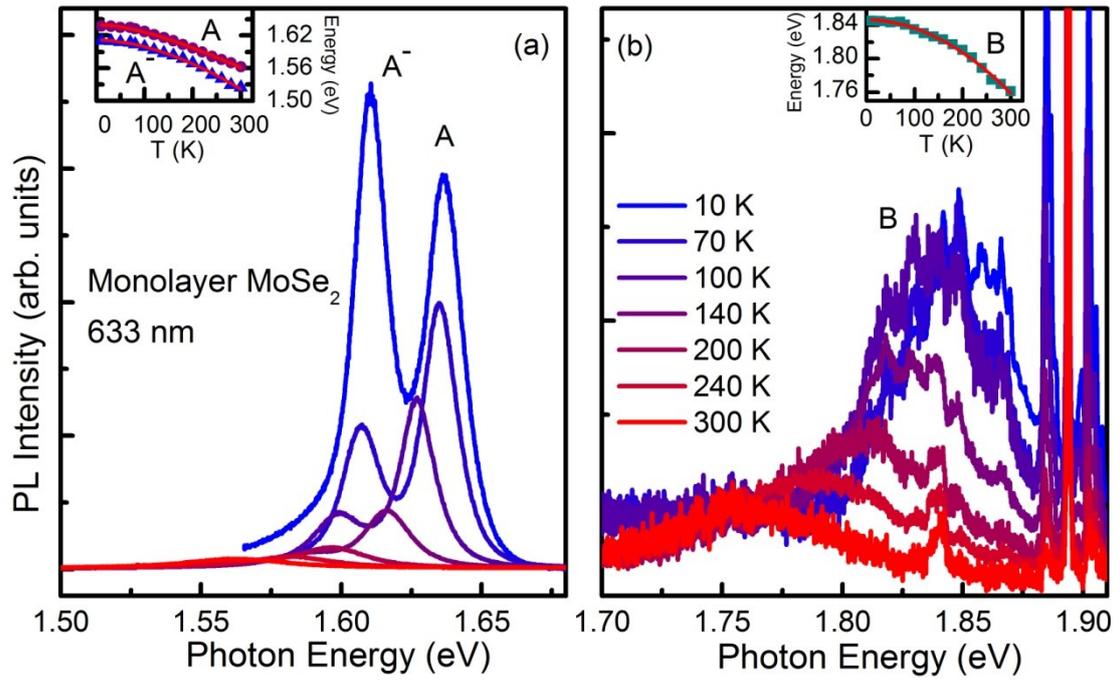

Fig.1

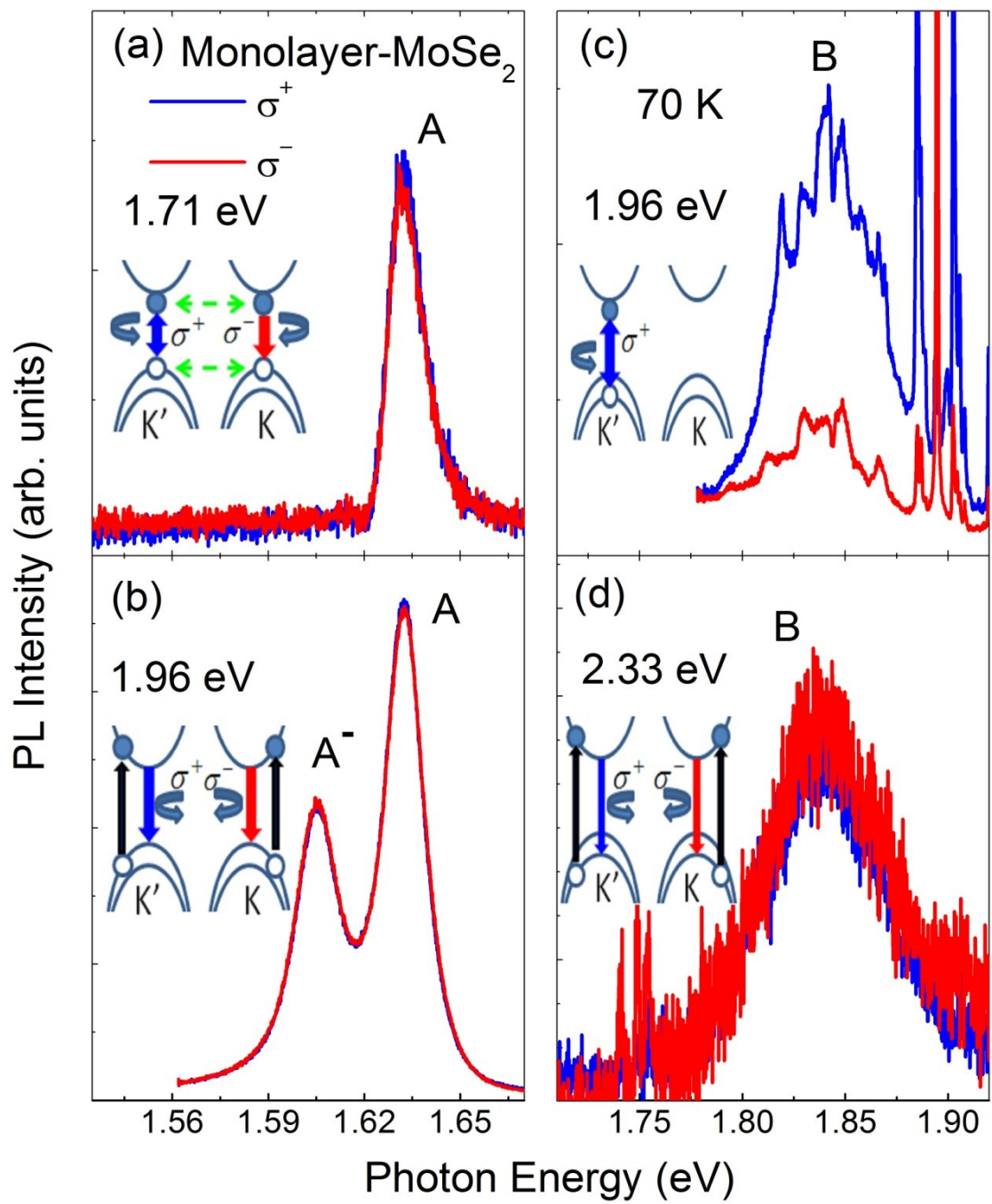

Fig.2

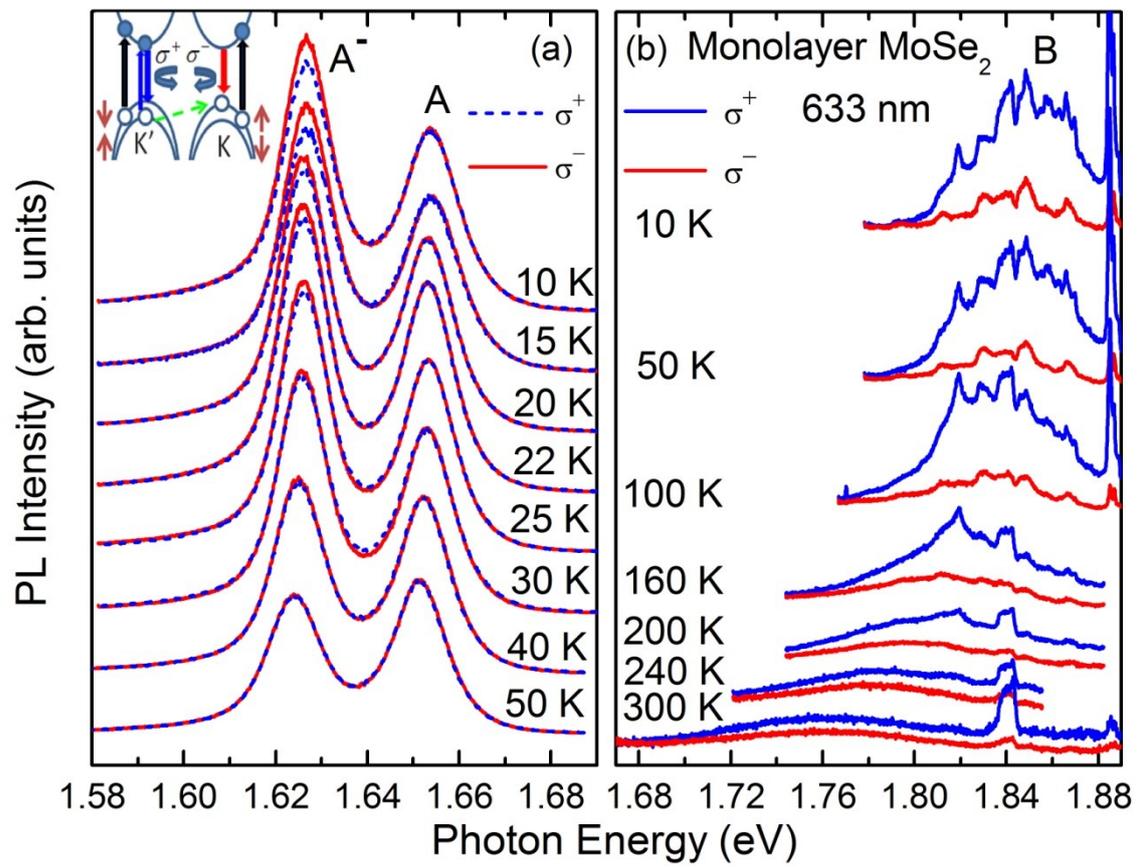

Fig. 3

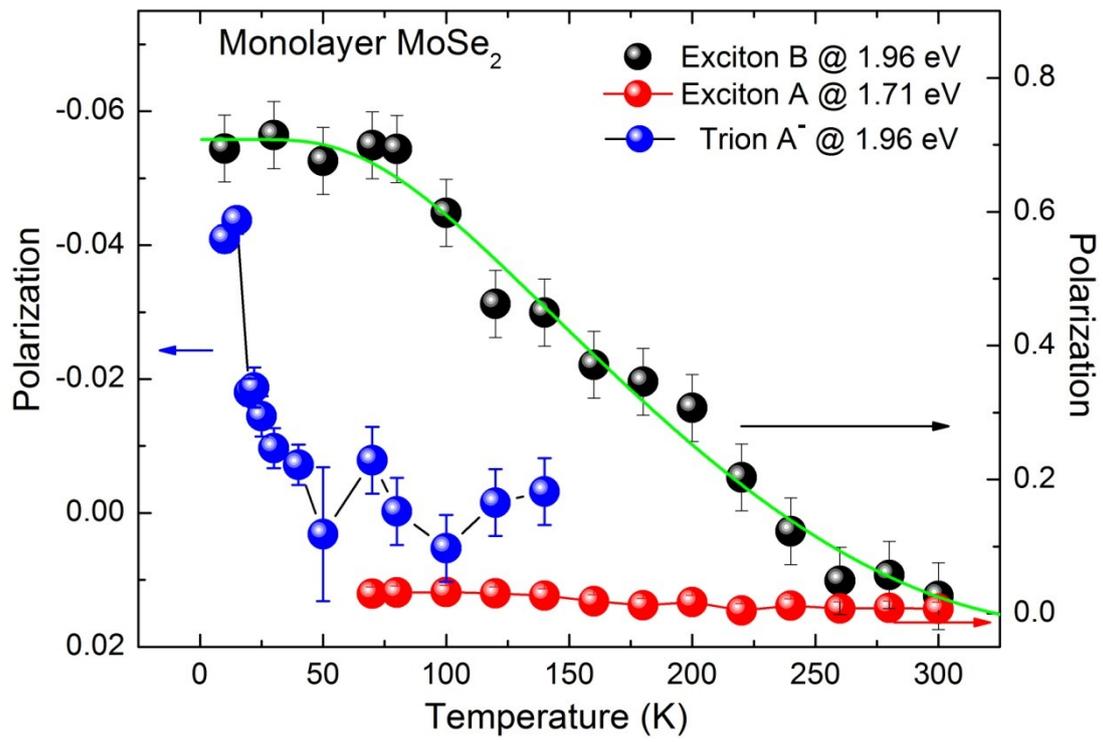

Fig. 4

Supplementary information

# Anomalous valley polarization in monolayer MoSe$_2$

Anmin Zhang,[1] Jiahe Fan,[1] Yusheng Li,[1] Jianting Ji,[1] Guihua Zhao,[1] Tianlong Xia,[1] Tengfei Yan,[2] Xinhui Zhang,[2] Wei Zhang,[1] Xiaoqun Wang[1] and Qingming Zhang[1]

[1]Department of Physics, Beijing Key Laboratory of Opto-electronic Functional Materials & Micro-nano Devices, Renmin University of China, Beijing 100872, P. R. China

[2]State Key Laboratory of Superlattices and Microstructures, Institute of Semiconductors Chinese Academy of Sciences, P.O. Box 912, Beijing 100083, P. R. China

Email: qmzhang@ruc.edu.cn

1. Sample characterizations
2. Polarized A-exciton PL spectra at low temperature
3. Temperature evolution of A-exciton PL peak

# 1. Sample characterizations

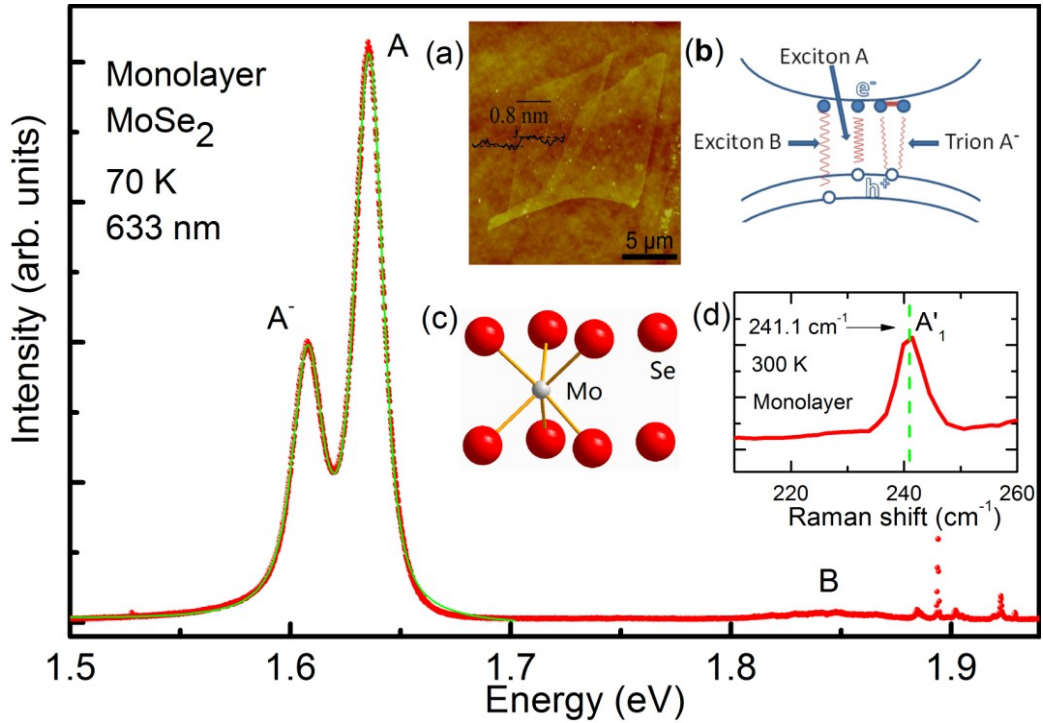

Fig. S1 PL spectra of monolayer $MoSe_2$ at 70 K. A/B and $A^-$ denote two kinds of excitons and trions respectively. The green fitting curve is obtained using two symmetric Voigt functions. Inset (a): AFM image determining one layer $MoSe_2$; Inset (b): microscopic hopping process for A/B excitons and $A^-$ trions; Inset (c): unit cell of monolayer $MoSe_2$; (d): $A_1'$ Raman phonon at room temperature.

Fig. S1 shows PL spectra under 633 nm excitation at 70 K. The strong peaks at 1.61 eV and 1.63 eV correspond to trions (A-) and A excitons, respectively. And the weaker PL peak of B excitons can also be seen at 1.84 eV. The sharp peaks around 1.9 eV are phonon excitations boosted under near-resonance conditions and silicon substrate mode (~520 $cm^{-1}$). PL spectra pumped by 633 nm laser at different temperatures are shown Fig. 1. At low temperatures, $A^-$ PL peak intensity is even higher than that of A excitons. But it rapidly decreases with increasing temperatures and is much lower than PL intensity of A excitons above 100 K, which is attributed to the electron escape from trions caused by thermal fluctuations.[1] The temperature dependence of A and $A^-$ peak positions given by symmetric Voigt fittings, perfectly follows Varshni empirical relationship (insets of Fig. 1),[2] indicating that electron/exciton-phonon coupling gives rise to the energy shift. By comparison of A and $A^-$ energies, we can further derive the trion binding energy of 28.4 meV at low temperatures.

## 2. Polarized A-exciton PL spectra at low temperature

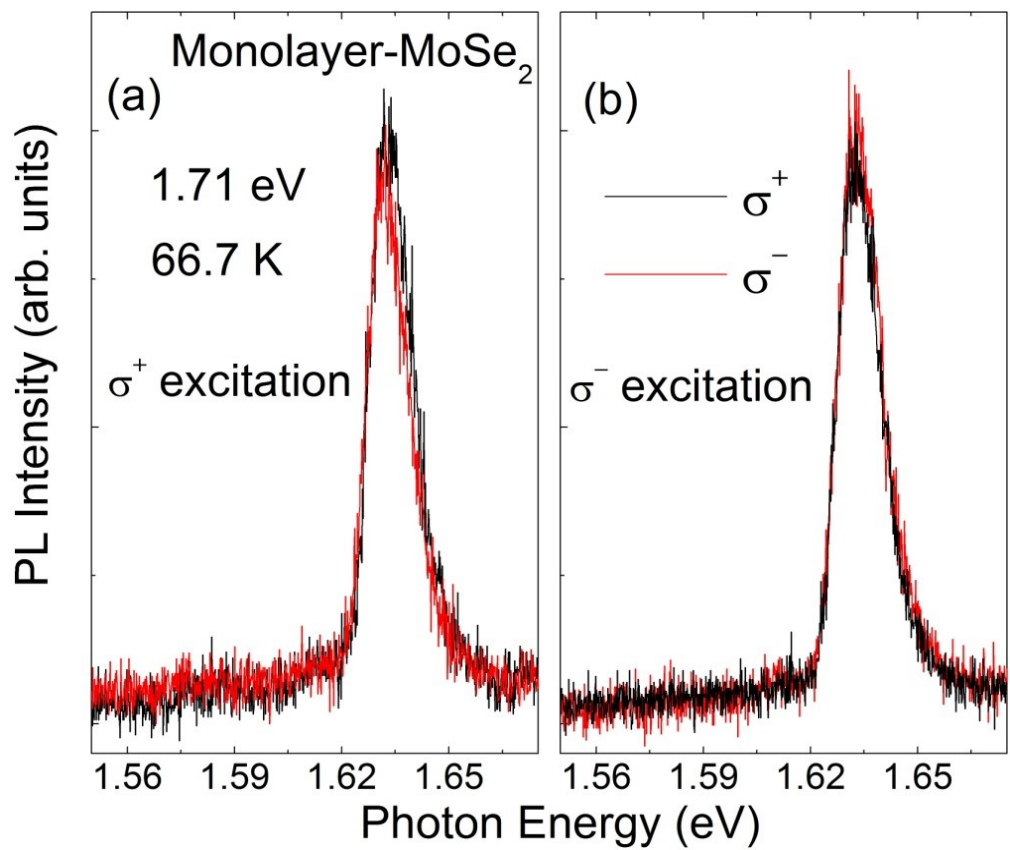

Fig. S2 Polarized PL spectra of monolayer MoSe$_2$ at low temperatures under near-resonance excitations. (a) Right-circularly polarized excitation light ($\sigma^+$); (b) Left-circularly polarized excitation light ($\sigma^-$).

Spin-valley coupling requires that the electrons at K' points can be excited only by right-circularly polarized light and their recombination only gives rise to right-circularly polarized PL spectrum. Similarly, both the excitation light and emitted PL are associated with the electrons at K points and required to be left-circularly polarized. However, Fig. S3 shows that PL is depolarized almost completely (>97%) though circularly polarized excitation light is used, indicating that strong inter-valley scattering eliminates the population difference of A excitons between K and K' points and eventually causes a complete depolarization of PL, as discussed in the context. Furthermore, the high similarity between (a) and (b) suggests a symmetric inter-valley scattering and confirms that the depolarization originates from the inter-valley scattering.

3. Temperature evolution of A-exciton PL peak

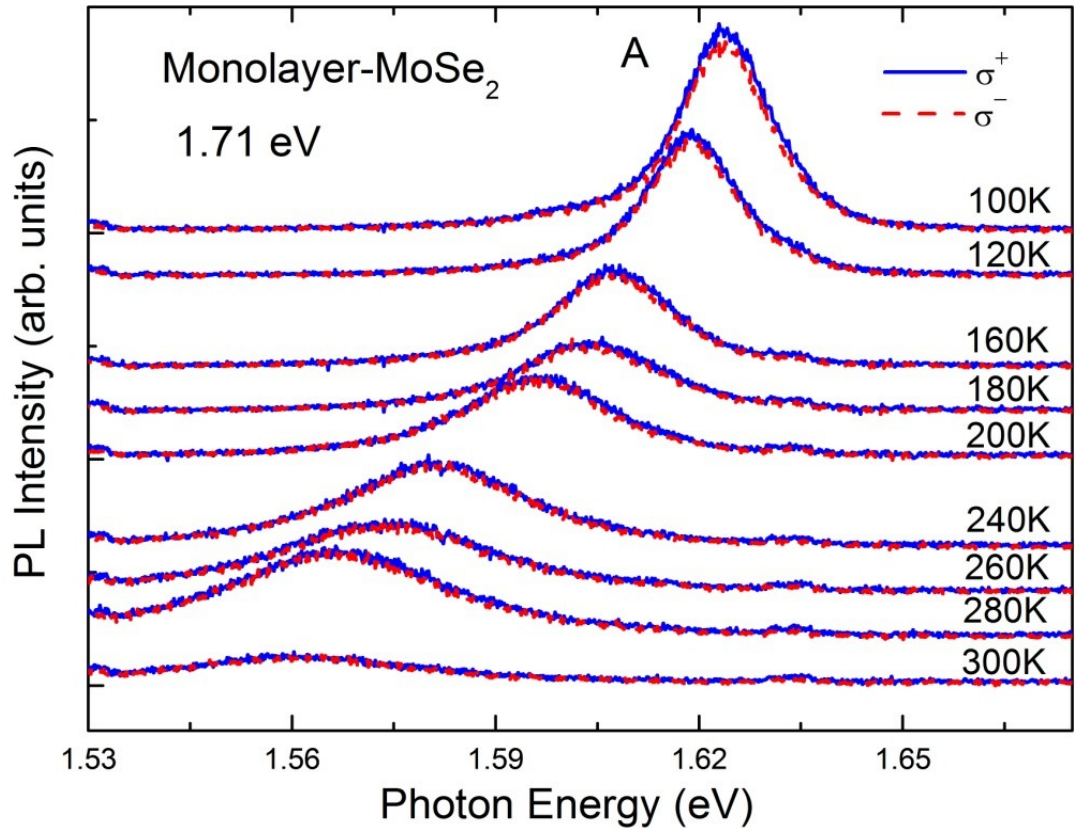

Fig. S3 Evolution of polarized PL with temperatures using right-circular excitation light. Under near-resonance condition, no clear valley polarization is observed with varying temperatures and PL peak is thermally broadened with increasing temperatures.

References

1. Jones, A.M. et al. Optical generation of excitonic valley coherence in monolayer WSe$_2$. Nature Nanotech. 8, 634–638 (2013).
2. Varshni, Y. P. Temperature dependence of the energy gap in semiconductors. Physica 34 149-154 (1967).